\documentclass[aps, prb, 10pt, twocolumn, showpacs, preprintnumbers,superscriptaddress, amsmath, amssymb]{revtex4-2}
\usepackage[pdftex]{graphicx}
\usepackage[utf8]{inputenc}
\usepackage[ngerman,english]{babel}
\usepackage{natbib}
\usepackage{color}
\usepackage{siunitx}
\usepackage{calc}
\usepackage{lipsum}
\usepackage[colorlinks=true, pdfstartview=FitV, linkcolor=blue,
citecolor=blue, urlcolor=blue]{hyperref}

\definecolor{green2}{rgb}{.0, .58, 0}

\newcount\colveccount
\newcommand*\colvec[1]{
        \global\colveccount#1
        \begin{pmatrix}
        \colvecnext
}
\def\colvecnext#1{
        #1
        \global\advance\colveccount-1
        \ifnum\colveccount>0
                \\
                \expandafter\colvecnext
        \else
                \end{pmatrix}
        \fi
}

\newcommand*\diff{\mathop{}\!\mathrm{d}}

\begin{document}
\title{Magnetic anisotropy of individual maghemite mesocrystals}
\author{B.~Gross} \affiliation{Department of Physics, University of
  Basel, 4056 Basel, Switzerland} \author{S.~Philipp}
\affiliation{Department of Physics, University of Basel, 4056 Basel,
  Switzerland} \author{E.~Josten} \affiliation{Ernst Ruska-Centre for
  Microscopy and Spectroscopy with Electrons (ER-C) and Peter Gr\"unberg
  Institute (PGI), Forschungszentrum J\"ulich, 52425 J\"ulich, Germany}
\author{J.~Leliaert} \affiliation{Department of Solid State Sciences,
  Ghent University, 9000 Ghent, Belgium} \author{E.~Wetterskog}
\affiliation{Department of Engineering Sciences, {\AA}ngstr\"om
  Laboratory, Uppsala University, 751 21 Uppsala, Sweden}
\author{L.~Bergstr\"om} \affiliation{Department of Materials and
  Environmental Chemistry, Stockholm University, 10691, Stockholm,
  Sweden} \author{M.~Poggio} \affiliation{Department of Physics,
  University of Basel, 4056 Basel, Switzerland}

\begin{abstract}
  Interest in creating magnetic metamaterials has led to methods for
  growing superstructures of magnetic nanoparticles. Mesoscopic
  crystals of maghemite ($\gamma\text{-Fe}_2\text{O}_3$) nanoparticles
  can be arranged into highly ordered body-centered tetragonal
  lattices of up to a few micrometers. Although measurements on
  disordered ensembles have been carried out, determining the magnetic
  properties of individual mesoscopic crystals is challenging due to
  their small total magnetic moment. Here, we overcome these
  challenges by utilizing sensitive dynamic cantilever magnetometry to
  study individual micrometer-sized $\gamma\text{-Fe}_2\text{O}_3$
  mesocrystals.  These measurements reveal an unambiguous cubic
  anisotropy, resulting from the crystalline anisotropy of the
  constituent maghemite nanoparticles and their alignment within the
  mesoscopic lattice.  The signatures of anisotropy and its orgins
  come to light because we combine the self-assembly of highly ordered
  mesocrystals with the ability to resolve their individual magnetism.
  This combination is promising for future studies of the magnetic
  anisotropy of other nanoparticles, which are too small to
  investigate individually.
\end{abstract}
\maketitle
\section{Introduction}
\label{sec:intro}
Maghemite ($\gamma$-Fe$_2$O$_3$) nanoparticles have a long history in magnetic recording applications~\cite{dronskowski_little_2001} and recent interest has been building for new technical and biomedical applications~\cite{shokrollahi_review_2017, tong_magnetic_2019}.  If small enough, these nanoparticles have been shown to be
superparamagnetic at room temperature~\cite{bedanta_supermagnetism_2008}. A number of studies have been carried out on superparamagnetic maghemite
nanoparticles, focusing on different physical aspects, including morphology~\cite{herlitschke_spin_2016, disch_spin_2014}, spatial magnetization distribution, shape anisotropy, spin disorder, superparamagnetic relaxation, and surface spin canting \cite{herlitschke_spin_2016, disch_spin_2014,  disch_quantitative_2012,zakutna_field_2020}.
In such measurements, which are all done on ensembles of nanoparticles, inter-particle interactions~\cite{andersson_interacting_2017} become relevant for small enough inter-particle distance.

At the same time, techniques to produce ordered assemblies of magnetic nanoparticles have also been developed~\cite{ahniyaz_magnetic_2007, josten_superlattice_2017, wetterskog_tuning_2016, wetterskog_colossal_2018, wetterskog_precise_2014,   disch_structural_2013, disch_shape_2011,sturm_nee_rosseeva_mesocrystals_2017,song_mesocrystalsordered_2010}. In fact, maghemite nanoparticles can now be self-assembled into highly ordered three-dimensional (3D) superlattice structures up to micrometers in size.  
Magnetic measurements have mostly been carried out on large ensembles
of mesocrystals, because conventional magnetometry techniques lack the
sensitivity to resolve the magnetic moment of an individual
mesocrystal.  DC and AC superconducting quantum interference device magnetometry on large mesocrystal ensembles has provided
temperature-dependent susceptibility measurements, yielding
superparamagnetic blocking temperatures~\cite{wetterskog_tuning_2016,
  brunner_self-assembled_2017,yang_synthesis_2004,
  kostiainen_hierarchical_2011,faure_2d_2013, wilbs_macroscopic_2017,
  fu_field-induced_2016}.  Measurements of high- and low-temperature
hysteresis loops have allowed the determination of saturation
magnetizations~\cite{yang_synthesis_2004,
  kostiainen_hierarchical_2011, faure_2d_2013, fu_field-induced_2016},
coercive fields~\cite{brunner_self-assembled_2017,cheon_magnetic_2006,
  kostiainen_hierarchical_2011}, and have shown first indications of
magnetic anisotropy~\cite{disch_spin_2010, faure_2d_2013,
  wetterskog_colossal_2018}.  Nevertheless, ensembles of mesocrystals
have a distribution of size, shape, and orientation and -- depending
on the density -- may interact with each other.  These complications
can obscure the magnetic properties of the individual mesocrystals.
Although Okuda et al.~\cite{okuda_top-down_2017} reported on microwave
absorption experiments on an individual mesocrystal, a cubic
Fe$_3$O$_4$-ferritin crystal
$50 \times 50 \times \SI{40}{\micro\meter}$ in size, the results did
not yield a detailed picture of the magnetism of the mesocrystal in
question.

Here, we use sensitive dynamic cantilever magnetometry
(DCM)~\cite{rossel_active_1996,harris_integrated_1999,stipe_magnetic_2001,gross_dynamic_2016}
to investigate individual 3D maghemite mesocrystals.  The well-defined
orientational order of the magnetic nanoparticles in the mesocrystal
allows us to unambiguously identify the presence of cubic magnetic
anisotropy, attributed to the crystal structure of the individual
maghemite particles.  In order to analyze our measurements, we develop
a model that describes the DCM response of a paramagnet and
distinguishes between a ferro- and paramagnet.  We find proof of
superparamagnetic behavior down to a blocking temperature
$T_b^{\text{spm}} = \SI{133}{\kelvin}$ for three different
mesocrystals.  Furthermore, an exchange bias and frozen spin state
below \SI{90}{\kelvin} provide evidence for a disordered surface spin
layer on the individual maghemite nanoparticles.

\section{Measurement Technique}
\label{sec:technique}
In DCM, as shown in Fig.~\ref{fig:setup}, the sample under
investigation is attached to the end of a cantilever, which is driven
into self-oscillation at its resonance frequency $f$.  Changes in this
resonance frequency $\Delta f = f - f_0$ are measured as a function of
a uniform applied magnetic field $\mathbf{H}$, where $f_0$ is the
resonance frequency at $H = 0$ (a few kilohertz for the cantilevers
used here).  $\Delta f$ reveals the curvature of the magnetic system's
free energy $\mathcal{F}$ with respect to rotations about the
cantilever oscillation
axis~\cite{mehlin_stabilized_2015,gross_dynamic_2016,modic_resonant_2018,geirhos_macroscopic_2020}:
\begin{figure*}[t]
  \includegraphics[width=14.2cm]{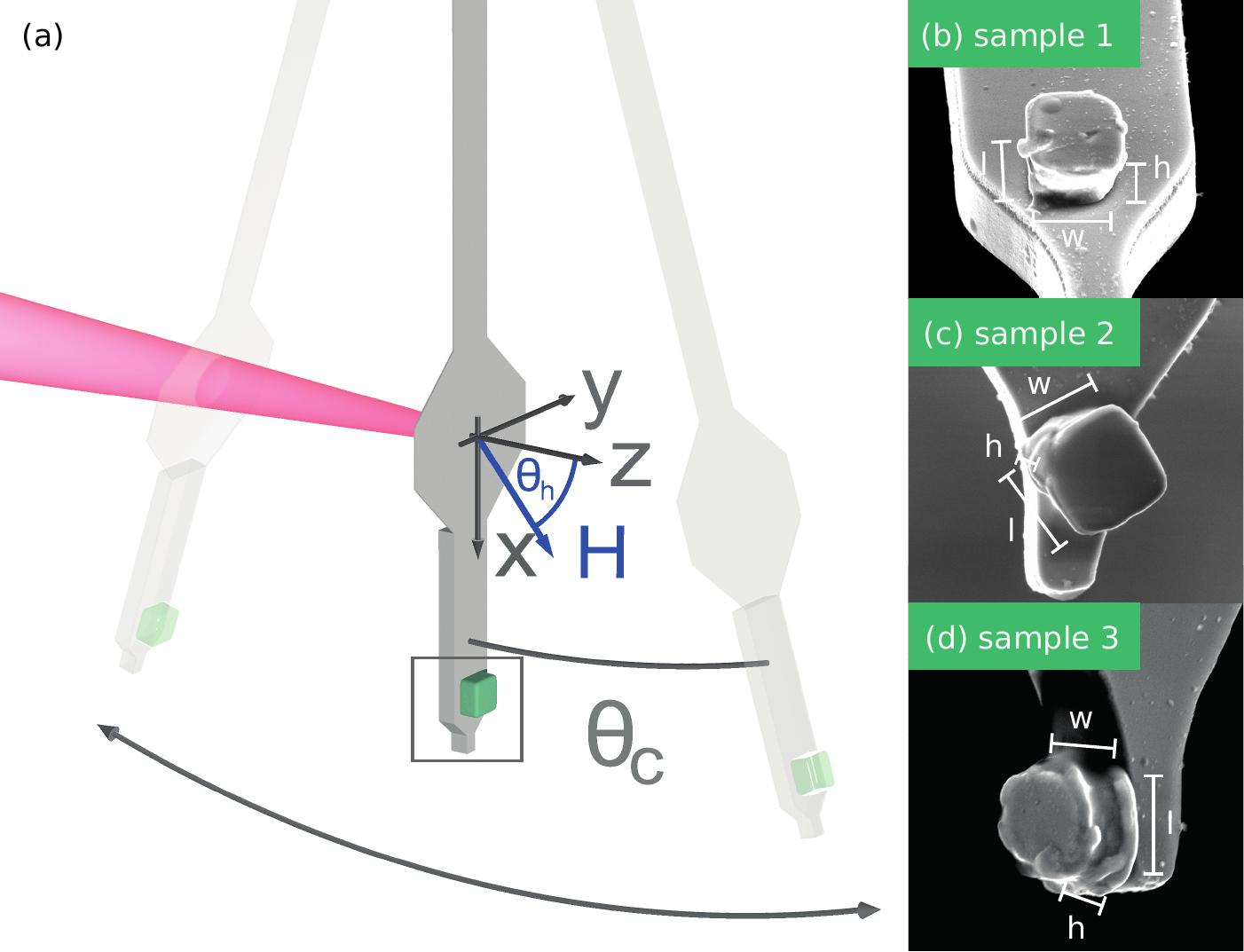}
  \caption{Sketch of the experiment and image of the samples. (a) An
    infrared laser, focused on the cantilever paddle, allows
    interferometric readout of the cantilever position. An externally
    applied, homogeneous magnetic field $\mathbf{H}$ can be rotated
    around the cantilever rotation axis $\mathbf{\hat{y}}$, spanning
    an angle $\theta_h$ with the $z$-axis. The magnetic specimen under
    investigation is shown as a green cube.  (b)-(d) SEM images of the
    investigated mesocrystals attached to cantilevers.  The length $l$
    $\times$ width $w$ $\times$ height $h$ of the samples are
    determined from the images to be
    $\SI{1.95}{\micro\meter} \times \SI{1.87}{\micro\meter} \times
    \SI{1.58}{\micro\meter}$,
    $\SI{1.7}{\micro\meter} \times \SI{1.89}{\micro\meter} \times
    \SI{1.12}{\micro\meter}$, and
    $\SI{1.55}{\micro\meter} \times \SI{1.54}{\micro\meter} \times
    \SI{1.35}{\micro\meter}$, for samples 1, 2, and 3, respectively.
    $h$ is defined to be along the [001] direction of the BCT crystal
    structure.}
\label{fig:setup}
\end{figure*}
\begin{equation}
  \Delta f = \frac{f_0}{2 k_0 l_e^2}\left( \left.\frac{\partial^2\mathcal{F}}{\partial\theta_c^2}\right|_{\theta_c=0}\right),
\label{eq:Df}
\end{equation}
where $k_0$ is the cantilever's spring constant, $l_e$ its effective
length, and $\theta_c$ its angle of oscillation.  Measurements of
$\Delta f$ are particularly useful for identifying magnetic phase
transitions~\cite{mehlin_stabilized_2015}, because it is discontinuous
for both first and second order phase
transitions~\cite{modic_resonant_2018}, just as the magnetic
susceptibility $\chi = \partial^2\mathcal{F}/\partial H^2$.
$\Delta f$ can also provide information on the switching, saturation
magnetization, coercivity, and the anisotropy of a magnetic system.

The ultrasensitive cantilevers are fabricated from undoped Si and are
\SI{90}{\micro\meter}-long, \SI{3.5}{\micro\meter}-wide and
\SI{0.1}{\micro\meter}-thick with a mass-loaded end and a
\SI{11}{\micro\meter}-wide paddle for optical detection, as shown in
Fig.~\ref{fig:setup}. The resonance frequency $f_0$ of the fundamental
mechanical mode used for magnetometry is between 5 and
\SI{6}{\kilo\hertz} with $k_0$ and $l_{e}$
$\SI{314}{\micro\newton/\meter}$ and \SI{74}{\micro\meter},
respectively. Mesocrystals are attached to the tips of ultrasensitive
Si cantilevers with nonmagnetic epoxy, using an optical microscope
equipped with precision micromanipulators.  The sample-loaded
cantilever is then mounted in a vibration-isolated closed-cycle
cryostat.  The pressure in the sample chamber is less than
$10^{-6}$~mbar and the temperature can be stabilized between 4 and
\SI{300}{\kelvin}.  Using an external rotatable superconducting
magnet, magnetic fields up to \SI{4.5}{\tesla} can be applied along
any direction spanning \SI{225}{\degree} in the plane of cantilever
oscillation ($xy$-plane), as shown in Fig.~\ref{fig:setup}.  This
direction is specified by $\theta_h$, which is the angle between
$\mathbf{H}$ and $\mathbf{\hat{z}}$, where $\mathbf{\hat{x}}$ is
parallel to the cantilever's long axis and $\mathbf{\hat{y}}$
coincides with its axis of oscillation.  The cantilever's flexural
motion is read out using a optical fiber interferometer employing
\SI{100}{\nano\watt} of laser light at
\SI{1550}{\nano\meter}~\cite{rugar_improved_1989}.  A piezoelectric
actuator mechanically drives the cantilever at $f_0$ with a constant
oscillation amplitude of a few tens of nanometers using a feedback
loop implemented by a field-programmable gate array.  This process
enables the fast and accurate extraction of $f_0$ from the cantilever
deflection signal.

\section{Samples}
\label{sec:samples}
\begin{figure}[t]
  \includegraphics[width=8.4cm]{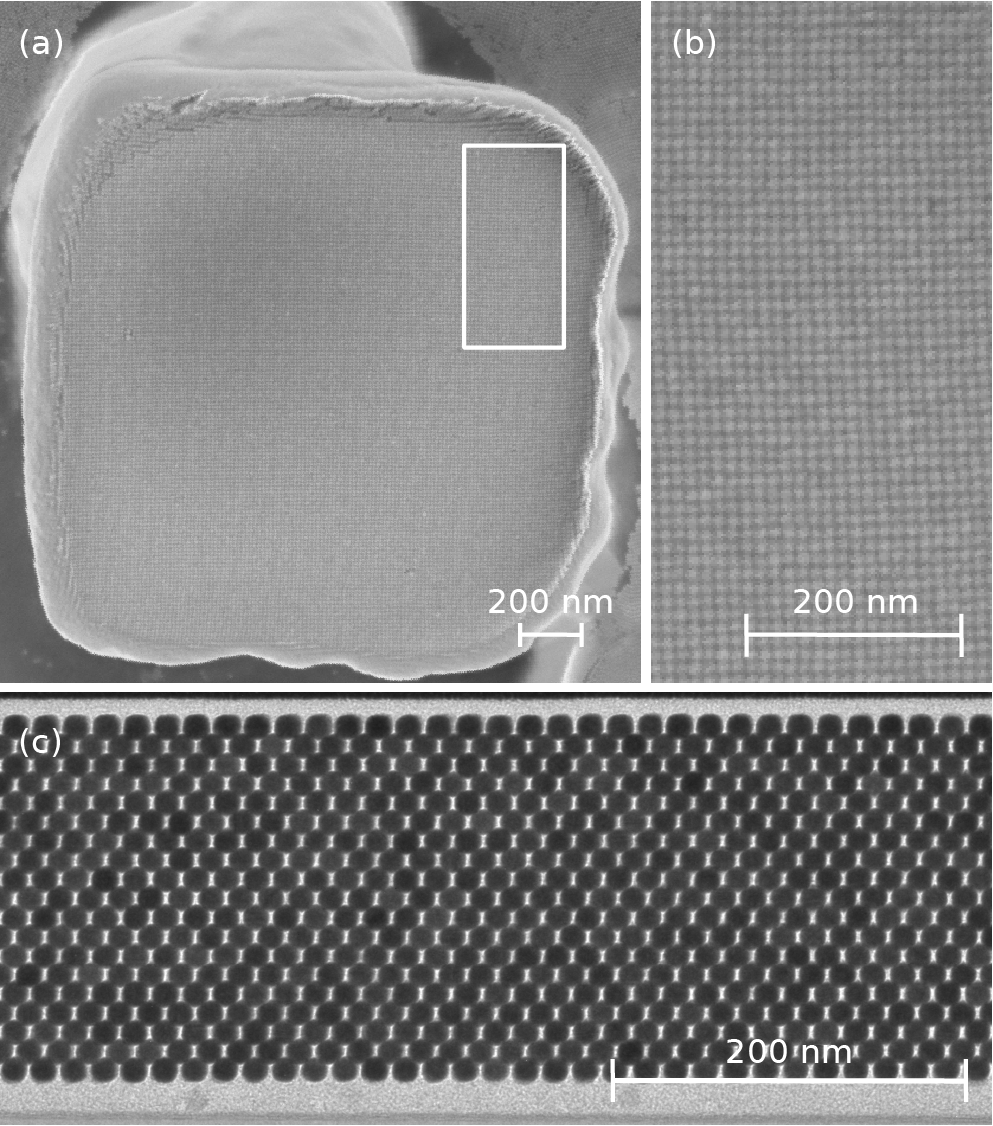}
  \caption{(a) SEM image of a mesocrystal showing the superlattice
    structure from a top view. (b) Zoom of region shown in (a). (c)
    Cross-sectional TEM image viewed from the [100] direction of a
    thinned mesocrystal layer deposited on a Si single crystal,
    showing the BCT structure.  Nanoparticles in the TEM image appear
    as dark circles.  }
\label{fig:images}
\end{figure}

The mesocrystal samples are composed of nanoparticles, which are
synthesized following a modified version of the metal oleate route
\cite{park_ultra-large-scale_2004, wetterskog_precise_2014}. These
particles consist of $\gamma$-Fe$_2$O$_3$ (maghemite) with less than
$\SI{10}{\percent}$ Fe$_3$O$_4$ (magnetite) content
\cite{disch_spin_2010, haggstrom_mossbauer_2008}.  The nanoparticles have an edge-length of $\SI{10.9}{\nano\meter}$, their atomic structure shows a crystalline inverse spinel structure, and their morphology can be described by a rounded cube model~\cite{josten_strong_2020}.
The micron-sized mesocrystals, i.e.\ 3D superlattices of the maghemite nanocubes arranged with a high degree of both positional and orientational order, have been carefully grown using an optimized evaporation-driven self-assembly process \cite{wetterskog_precise_2014, wetterskog_tuning_2016, josten_strong_2020}.
%
%
Small angle X-ray diffraction performed on an individual mesocrystals reveals a body centered tetragonal (BCT) crystal lattice with an in-plane lattice constant $a = \SI{13.47}{\nano\meter}$, and an out-of-plane lattice constant $c = \SI{15.08}{\nano\meter}$ \cite{josten_strong_2020},
cf. Fig.~\ref{fig:images} (c) for a cross-sectional transmission electron microscopy (TEM) image of a thinned mesocrystal layer showing the BCT structure.  
The self-assembly process for the mesocrystals is both size- and shape-selective~\cite{wetterskog_tuning_2016, josten_strong_2020}, i.e.\ the mesocrystals are composed of particles with a size dispersity which is drastically smaller than that in the initial dispersion~\cite{josten_strong_2020}.
Fig.~\ref{fig:images} (a) and (b) show scanning electron microscopy (SEM) images of a typical mesocrystal.
\section{Results and Analysis}
\label{sec:results}
We investigate three different mesocrystals, which are each attached
in different orientations to the end of a cantilever, as shown
Fig.~\ref{fig:setup}. The different orientations allow us to probe the
anisotropy in different planes of the superlattice. The crystals have
slightly different sizes, which we estimate from SEM images and list
in the caption of Fig.~\ref{fig:setup}.

DCM experiments are first carried out at $T = \SI{270}{\kelvin}$.
\begin{figure}[t]
  \includegraphics[width=8.28cm]{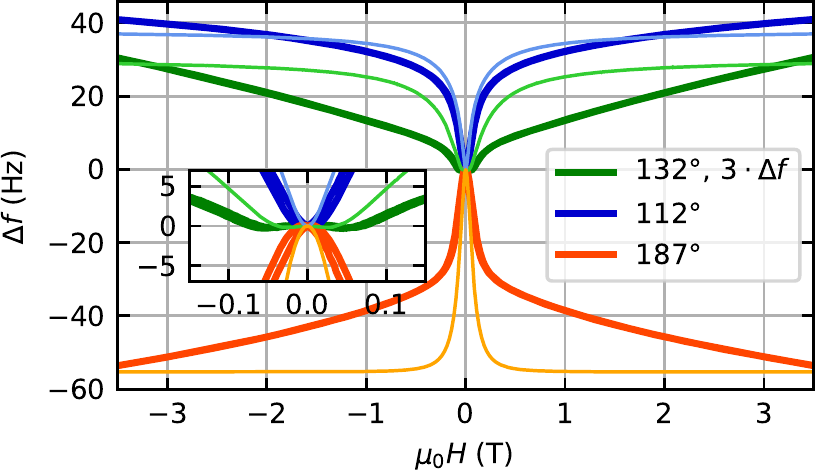}
  \caption{DCM measurements at \SI{270}{\kelvin}.  Dark thick lines
    show measured $\Delta f(H)$ in sample 1 for orientations
    $\theta_h = \SI{112}{\degree}$, $\SI{187}{\degree}$ (blue and
    orange) and $\Delta f(H)$ in sample 3 for
    $\theta_h = \SI{132}{\degree}$ (green). Light thin lines with
    similar coloring show the corresponding simulations according to a
    model for a thermally activated Stoner-Wohlfarth particle.  The
    inset shows an enlargement around $H=0$.}
\label{fig:HT2}
\end{figure}
In Fig.~\ref{fig:HT2}, we plot $\Delta f(H)$ measured in samples 1 and
3, where the external field is swept between $H = \pm\SI{3.5}{T}$ for
three orientations of $\theta_h$.  
Most DCM curves show a V- or \reflectbox{\rotatebox[origin=c]{180}{V}}-shape,
depending on the orientation of the applied field.  At low field, well
below magnetic saturation, some curves present a W-shape, as seen in
the inset.  In this regime, $\Delta f(H)$ shows a small hysteresis with a coercive field of $\mu_0 H_c\approx\SI{10}{\milli\tesla}$ for all three mesocrystals.
At high fields, the curves show an asymptotic behavior and approach either a positive or negative
$\Delta f$, depending on the orientation of the external field.  The
presence of this asymptotic frequency shift implies a magnetic anisotropy in the system
\cite{gross_dynamic_2016}.  A positive (negative) asymptote signifies
alignment with a magnetically easy (hard) direction.   We further investigate the
mesocrystals' magnetic anisotropy by measuring the DCM frequency shift
of the 3 samples as a function of applied field angle $\theta_h$ for
$\Delta f(H = \SI{3.5}{\tesla}/\mu_0, \theta_h)$, at which the samples
are near magnetic saturation.  The bulk value of the saturation
magnetization $M_s$ is approximately
$3 \times 10^5\si{\ampere/\meter} =
\SI{0.38}{\tesla}/\mu_0$~\cite{disch_spin_2014}.  Fig.~\ref{fig:HT3}
shows polar plots of these measurements, which show evidence of
multi-axial anisotropy, most clearly in the case of sample 3.
%
%
\subsection{High-field Limit}
\label{subsec:high-field}
\begin{figure*}[t]
\includegraphics[width=16.9cm]{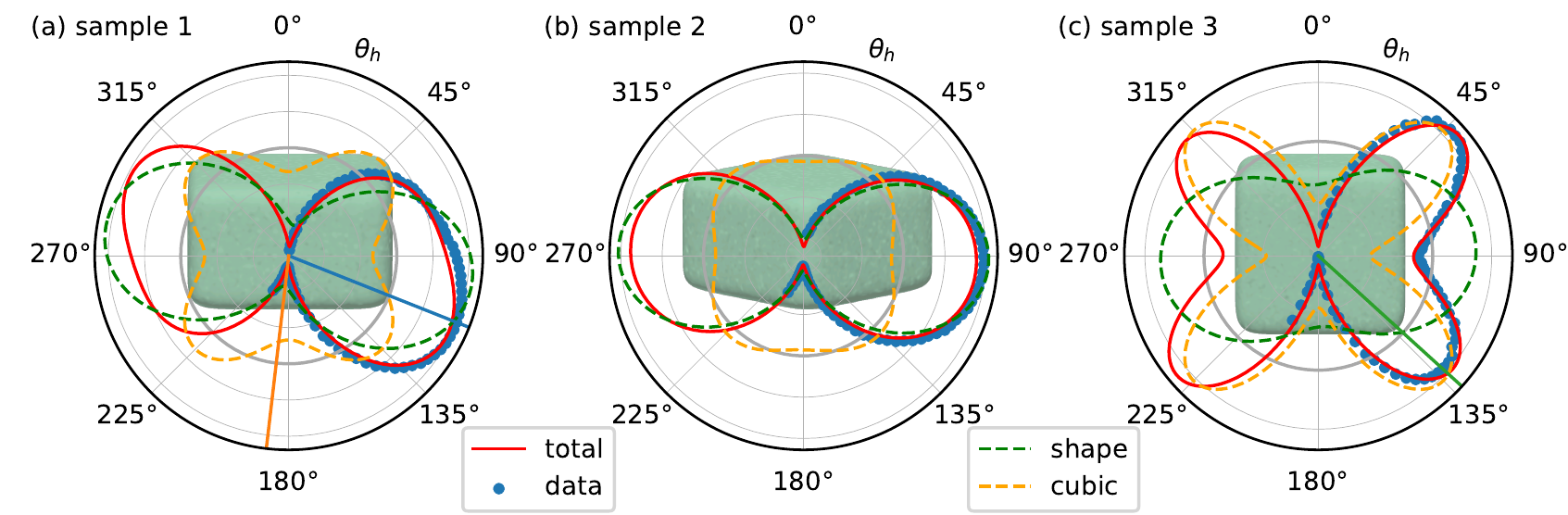}
\caption{Angular dependence of DCM in the high-field limit.  (a)-(c)
  Polar plots of the high-field frequency shift
  $\Delta f (H = \SI{3.5}{\tesla}/\mu_0, \theta_h)$ for sample 1 to 3
  (blue dots) and corresponding fits (red, dashed green, and dashed
  orange lines). Straight, radial lines indicate the direction of the
  measurements and calculations in Fig.~\ref{fig:HT2}, according to
  the color code. Gray polar lines indicate $\Delta f$ in steps of (a,
  b) $\SI{20}{\hertz}$ and (c) $\SI{10}{\hertz}$, with a thick line
  for $\Delta f=0$. In (a) a sketch of the mesocrystal is shown in the
  background to illustrate its orientation. }
\label{fig:HT3}
\end{figure*}

In order to understand these measurements, we consider the
contributions of the possible types of magnetic anisotropy.  For a
system with uniaxial anisotropy only, set by a uniaxial anisotropy
constant $K_{u1}$ (assuming higher order terms are negligible, e.g.\
$K_{u2} = 0$), the frequency shift in the limit of large applied
field, i.e.\ $H \gg |K_{u1}/ (\mu_0 M_s)|$, is given
by~\cite{gross_dynamic_2016}
%
%
\begin{equation}
\begin{split}
  \Delta f_{\text{uniaxial}} = &\frac{K_{u1} V f_0}{4 k_0 l_e^2} \bigg(\cos 2\theta_h\Big(1+3\cos 2\theta_h
		-2\cos2\phi_u \\& \sin^2\theta_u\Big)+4\cos\phi_u\sin 2\theta_h\sin 2\theta_u\bigg),
\end{split}
\label{eq:limit1}
\end{equation}
where $V$ is the volume of the magnetic object.  $\theta_u$, $\phi_u$
and $\theta_h$ denote the orientation of the anisotropy axis and the
external field, respectively.  Because the external field is
restricted to the $xz$-plane in our setup, no azimuthal angle of the
applied field appears in the equation.  Similar contributions can be
deduced for other types of anisotropy. For cubic anisotropy (again
assuming higher order terms are negligible, e.g.\ $K_{c2}=0$), in the
limit of large applied field, i.e.\ $H \gg |K_{c1}/ (\mu_0 M_s)|$, we
find:
%
%
\begin{equation}
\begin{split}
  \Delta f_{\text{cubic}} = & -\frac{K_{c1} V f_0}{8 k_0 l_e^2}\cdot
		 \bigg(\cos2\theta_h + 7\cos4\theta_h\\
		& - 2\Big(1+2\cos2\theta_h\Big)\cos4\phi_u\sin^2\theta_h\bigg).
	\end{split}
\label{eq:limit2}
\end{equation}
In this case, one of the three anisotropy vectors is assumed to be
orthogonal to the cantilever plane
($\mathbf{\hat{u}_3}\parallel \mathbf{\hat{z}}$). Therefore,
$\mathbf{\hat{u}_1}$ and $\mathbf{\hat{u}_2}$ lie in the $xy$-plane,
and $\phi_u$ is the angle between $\mathbf{\hat{u}_1}$ and
$\mathbf{\hat{x}}$.  In light of these relations, we analyze the
measured dependence $\Delta f(H = \SI{3.5}{\tesla}/\mu_0, \theta_h)$
in Fig.~\ref{fig:HT3}.  To fit the measurements, we find that a sum of
a uniaxial, cf.\ eq.\ (\ref{eq:limit1}), and a cubic, cf.\ eq.\
(\ref{eq:limit2}), anisotropy is required.  A quantitative
determination of $K_{u1}$ and $K_{c1}$ is not possible from these
particular fits, because \SI{4.5}{\tesla}, which is the highest field
that we can apply in our apparatus, does not fully satisfy the
high-field limit.  We can, however, determine the relative weight of
the two contributions to the anisotropy.

%
%
%
The cubic shape of the nanoparticles within a mesocrystal and the fact
that random fluctuations in the shape of individual particles average
out over the entire mesocrystal allows us to neglect uniaxial
anisotropy due to the shape of individual nanoparticles.  Therefore,
we assume any observed uniaxial anisotropy to be the result of an
effective shape anisotropy of the mesocrystal as a whole. The dipolar
interactions between the nanoparticles and, therefore, the shape and
lattice spacing of the mesocrystal determine this effective shape
anisotropy, in analogy to the shape anisotropy of a continuous
magnetic solid~\cite{aharoni_introduction_2000,
  skomski_permanent_1999}.  A 1D equivalent of such an effective
anisotropy has previously been used to describe chains of iron oxide
nanoparticles~\cite{wetterskog_colossal_2018,
  phatak_magnetostatics_2011}. Micromagnetic simulations show that
this effective magnetic shape anisotropy is dominated by the overall
shape of the mesocrystal samples, rather than by the small difference
between the in- and out-of-plane lattice spacings of their BCT lattice
(cf. Appendix~\ref{app:BCT}).

A cubic component of the anisotropy may be present in the mesocrystals
as a result of the cubic shape of the overall mesocrystal or of the
individual maghemite nanoparticles~\cite{chekanova_micro_2015}, their
crystalline anisotropy~\cite{birks_properties_1950,
  takei_vacancy_1966}, or their surface anisotropy, as suggested in
Refs.~\citenum{kachkachi_surface-induced_2006,
  garanin_surface_2003,yanes_effective_2007}.  The latter is
calculated to be relevant only for particles with up to about 100
atoms per dimension, which is exceeded by our nanoparticles.
Micromagnetic simulations show that the contribution from the cubic
shape of the mesocrystals and the constituent nanoparticles are both
at least one order of magnitude too small to account for the
anisotropy observed in our DCM measurements
(cf. Appendix~\ref{app:cubic}).  On the other hand, the crystalline
contribution should appear in our measurements, due to the alignment
of the individual nanoparticles with respect to each other.

Indeed, in all three mesocrystal orientations shown in
Fig.~\ref{fig:HT3}, both the uniaxial and the cubic components of the
fitted curves match the expected orientation of the mesocrystal and
its constituent crystalline nanoparticles.  Furthermore, the magnitude
of the uniaxial term is seen to scale with the overall shape of the
mesocrystals: e.g.\ sample 3, the most symmetric mesocrystal
(cf. caption of Fig.~\ref{fig:images}), shows a nearly vanishing
uniaxial anisotropy. 
%
%
%
\subsection{Full Field Dependence}
\label{subsec:FullDep}
In order to extract quantitative values for the uniaxial and cubic
anisotropies and to understand the full field dependence of the
measured $\Delta f (H)$ curves shown in Fig.~\ref{fig:HT2}, including
the low-field regime, we develop a model of the entire DCM response.
Although such models are well-established for ferromagnetic systems,
through a simple Stoner-Wohlfarth approach \cite{gross_dynamic_2016}
or using micromagnetic simulations \cite{gross_dynamic_2016,
  mehlin_observation_2018, rossi_magnetic_2019}, no such framework
exists for superparamagnetic systems.  
At $T = \SI{270}{\kelvin}$
we expect individual maghemite cubes of the given dimensions to be superparamagnetic, despite being embedded in a superlattice structure~\cite{wetterskog_colossal_2018}.
The dense arrangement of the nanoparticles within a mesocrystal involves significant inter-particle dipolar interactions. These interactions alter the
magnetic behavior of the system, resulting in a shifted blocking temperature, hysteresis, or even a suppression of superparamagnetism. Below the blocking temperature, we expect the system to become ferromagnetic.

We follow Ref.~\citenum{garcia-palacios_statics_2007} to develop a DCM
model for the simplest superparamagnet: an individual, thermally
activated Stoner-Wohlfarth particle. To do so, an effective single
particle Hamiltonian is constructed, and the corresponding partition
function and free energy $\mathcal{F}$ are numerically
calculated. Then, thermodynamic quantities are extracted by
determining the corresponding derivatives, as shown in
Appendix~\ref{app:model}. In the case of DCM, $\Delta f$ can be found
according to eq.\ (\ref{eq:Df}).  In Appendix~\ref{app:modelex}, we
verify the applicability of this model by recovering the ferromagnetic
Stoner-Wohlfarth response, as calculated in
Ref.~\citenum{gross_dynamic_2016}, for a thermally activated particle
with only magnetic shape anisotropy in the low temperature limit.  In
this model, paramagnetic behavior sets in as the temperature is
increased.

For a mesocrystal of superparamagnetic nanoparticles, we then consider
$n$ interacting superparamagnets, where $n$ is the number of
nanoparticles in the mesocrystal lattice.  Each obeys an effective
Hamiltonian, which includes both uniaxial and cubic magnetic
anisotropy terms.  The uniaxial term reflects
the dipolar interactions between particles, i.e.\ the effective shape
anisotropy of the whole mesocrystal, and the cubic term reflects the
crystalline anisotropy of each particle.  Because we model the
inter-particle interaction with an effective single particle term in
the Hamiltonian, the model has limited validity, similar to approaches
relying on mean-field Hamiltonians~\cite{atherton_mean_1990}.  Model
parameters for each mesocrystal are summarized in
Appendix~\ref{app:model}.  $n$ is estimated from the mesocrystal
dimensions, as determined from SEM images and adjusted to match the
measurements at high field. To account for the expected presence of a
disordered surface spin layer and to adequately fit the data, we model
the individual maghemite particles to be slightly smaller, $9$ rather
than $\SI{10.9}{\nano\meter}$ on a side.  $M_s$ is taken to be
$3 \times 10^5\si{\ampere/\meter}$~\cite{disch_spin_2014}.  We use the
same magnitude of $K_{c1}$ for all mesocrystals, since this term
represents the crystalline anisotropy of the individual maghemite
nanoparticles, and set it to $\SI{-3.0}{\kilo\joule/\meter^3}$ to best
match the measurements.  This value is smaller than the
$\SI{-4.7}{\kilo\joule/\meter^3}$ of bulk
$\gamma\text{-Fe}_2\text{O}_3$, perhaps due to interparticle
interactions~\cite{rebbouh_57mathrmfe_2007}.  Fitting the data also
yields $K_{u1} = 9.7$, $20.8$, and $\SI{2.1}{\kilo\joule/\meter^3}$ or
$D_u =-0.17$, $-0.37$, and $-0.04$
in terms of the effective demagnetization factor for samples 1, 2, and 3,
respectively.  All of these values should be treated as approximate,
given the simpleness of the model and the uncertainty (up to
$20\%$) in precisely determining the magnetic volume of the
mesocrystal samples.

Calculated $\Delta
f(H)$ curves are plotted along with measured $\Delta
f(H)$ curves for the same field orientations in
Fig.~\ref{fig:HT2}. The model adequately captures the overall features
of the measurements, including V-,
\reflectbox{\rotatebox[origin=c]{180}{V}}-, and W-shapes.  This
agreement is evidence that the particles making up the mesocrystals
are indeed in a superparamagnetic state at $T =
\SI{270}{\kelvin}$.  Most notably, the
\reflectbox{\rotatebox[origin=c]{180}{V}}-shape is observed for a hard
axis alignment of the external field (yellow curve) and is in strong
contrast to the signature of a ferromagnet
(cf. Appendix~\ref{app:modelex}). The model also explains the
occurrence of W-shaped curves: it is a consequence of the opposing
sign of the cubic crystalline and the uniaxial shape anisotropy
contributions for certain orientations of the external field, e.g. for
$\theta_h \approx
\SI{135}{\degree}$, cf. Fig.~\ref{fig:HT3} (c). A pure cubic system in
this orientation leads to a relatively broad V-shape, while a pure
uniaxial system leads to a relatively sharp
\reflectbox{\rotatebox[origin=c]{180}{V}}-shape with a small negative
high-field asymptote.  The presence of both anisotropies and their
resultant competition produces a W-shaped curve.

Despite this agreement, the model predicts high-field $\Delta f(H)$
asymptotes that saturate at lower field than in experiment, presumably
as a consequence of the interactions between the particles, which we
do not fully consider.  Furthermore, in contrast to the experiments,
the model does not predict hysteresis as a function of $H$. However,
introducing strong interactions (e.g.\ with a mean field approach) or
large anisotropies (increasing the anisotropy constants) to the model
leads to ferromagnetic behavior, which includes hysteresis.  We thus
hypothesize that the observed hysteresis originates from the
presence of the inter-particle interactions.
%
\subsection{Behavior at low temperature}
\label{subsec:lowtemp}
Temperature dependent measurements of the DCM response down to
$\SI{5}{\kelvin}$ allow us to extract further information about the
mesocrystals.  Measurements of $\Delta f (T)$ after a mesocrystal is
cooled in an applied field, i.e.\ field cooling (FC), and in zero
field, i.e.\ zero-field cooling (ZFC) show the onset of ferromagnetism
below a superparamagnetic blocking temperature
$T_b^{\text{spm}} \approx \SI{133}{\kelvin}$, as discussed in
Appendix~\ref{app:blockT}.  Furthermore, measurements of magnetic
hysteresis in $\Delta f(H)$ as a function of decreasing temperature
show that a more complicated magnetic state emerges at lower
temperatures.  In particular, measurements show an exchange bias field
$H_{\text{eb}}$ emerging below a blocking temperature of
$T_{b}^{\text{eb}}\approx\SI{90}{\kelvin}$.  Just below
$\SI{50}{\kelvin}$ the coercivity $H_c$ is seen to increase
dramatically, possibly indicating a magnetic transition of unknown
origin.

\begin{figure}[t]
\includegraphics[width=7.07cm]{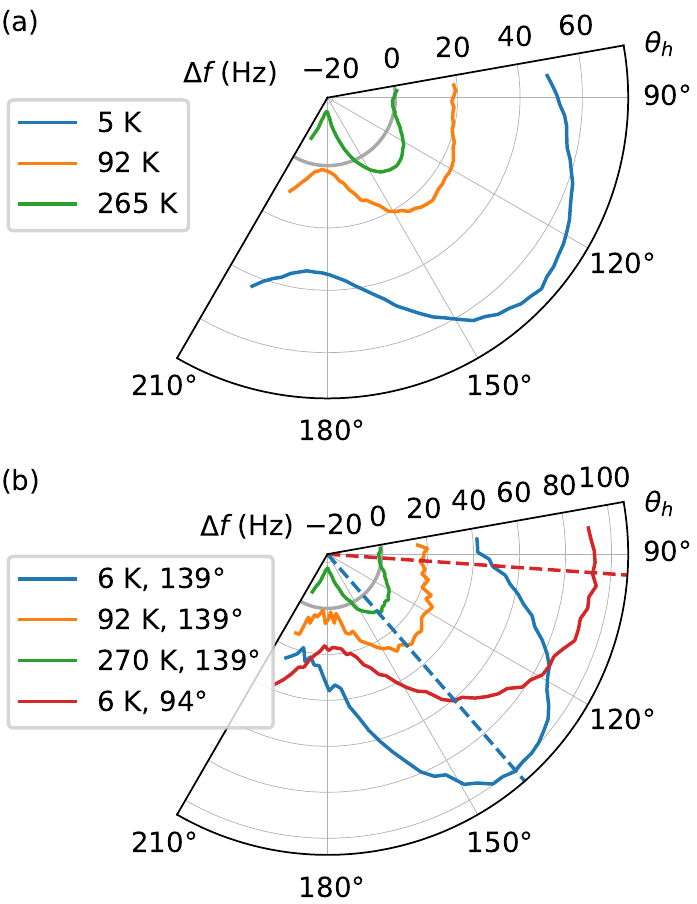}
\caption{High-field frequency shift
  $\Delta f (H = \SI{3.5}{\tesla}/\mu_0, \theta_h)$ for sample 3
  depending on the external field direction for various
  temperatures. The sample undergoes (a) ZFC and (b) FC. The magnetic
  field for FC is applied at the angle indicated by the color-coded
  dashed line.}
\label{fig:TD4}
\end{figure}
Fig.~\ref{fig:TD4} shows high-field DCM data, similar to that shown in
Fig.~\ref{fig:HT3}~(a)-(c), measured at different temperatures under
both ZFC and FC.  The measurements show that under ZFC, shown in
Fig.~\ref{fig:TD4}~(a), while the shape of $\Delta f (\theta_h)$ is
preserved down to $\SI{5}{\kelvin}$, its magnitude increases with
decreasing temperature.  This behavior indicates an increase of either
the anisotropy, the saturation magnetization, or both.  Measurements
under FC give similar results as those under ZFC regardless of the FC
field direction for all but the lowest temperature measurements.
Below \SI{90}{\kelvin}, however, the shape, orientation, and magnitude
of the signal change for ZFC and FC.  The direction of the maximum in
$\Delta f (\theta_h)$ is observed to follow the direction of the FC
field.  This reorientation of the magnetic anisotropy by the FC field
is observed both when the sample is cooled in an external field
applied at $\SI{139}{\degree}$ and $\SI{94}{\degree}$, as shown in
Fig.~\ref{fig:TD4}~(b).  Upon heating above $\SI{90}{\kelvin}$, the
original shape and magnitude of $\Delta f (\theta_h)$ is restored.

Similar observations have been made in dilute systems of particles,
where inter-particle interactions are
negligible~\cite{martinez_low_1998, andersson_interacting_2017}.  As
in those and similar studies~\cite{ramos-guivar_superspinglass_2019,
  shendruk_effect_2007, phan_exchange_2016}, including small angle
neutron scattering measurements~\cite{disch_quantitative_2012,
  zakutna_field_2020}, we conclude that the individual nanoparticles
are likely surrounded by a disordered system of surface spins. Below
$T_{b}^{\text{eb}}$, these spins freeze, leading to the observed
exchange bias and an additional magnetic anisotropy that can be set
and oriented by FC.

Given the agreement of these previous measurements with our data, we
find that frozen surface spins are more likely to explain the observed
exchange bias and anisotropy than frustration of the core spins in our
densely-packed superlattice of nanoparticles.  Nevertheless, the
configuration of the core spins at low temperatures remains
unknown. States such as superferromagnetic and
superantiferromagnetic~\cite{bedanta_supermagnetism_2008} ordering or
a superspin glass~\cite{jonsson_aging_1995} are potentially
present. Further experiments, such as real space imaging or aging
experiments are necessary to pin down the mesocrystal's
low-temperature magnetic configuration. The kink in the temperature
dependence of $H_c$ around $\SI{50}{\kelvin}$, which is discussed in
Appendix~\ref{app:belowBlock}, may be an indication of a phase
transition of such a superspin state.

\section{Conclusion}
\label{sec:conclusion}
%
%
In conclusion, our measurements reveal the different contributions to the magnetic anisotropy of a mesocrystal of maghemite nanoparticles, most notably a cubic component, which we attribute to the crystalline anisotropy of the constituent nanoparticles.  
A model considering interacting superparamagnetic nanoparticles captures most of our findings.  The system remains in a superparamagnetic state down to
$T_b^{\text{spm}} \approx \SI{133}{\kelvin}$.  Below $T_b^{\text{eb}} \approx \SI{90}{\kelvin}$, exchange bias and a frozen spin state are present in the system, consistent with a disordered layer of surface spins on the individual nanoparticles, as observed in earlier works.

We emphasize that the observation of cubic magnetic anisotropy in
these nanoparticles is only possible, because of the combination of
two techniques: the size-selective self-assembly of nanoparticle mesocrystals with a narrow size distribution and a high degree of
orientational order~\cite{wetterskog_tuning_2016, josten_strong_2020}, and measurement by DCM, which is sensitive enough to resolve the
magnetism of individual mesocrystals.  This ability to isolate the
magnetic response of a single mesocrystal overcomes the limitations of
measuring ensembles, which are composed of mesocrystals of varying
size, shape, and orientation.  This disorder and the potential for
interactions between mesocyrstals complicates the determination of
their individual magnetic properties and those of their constituent
nanoparticles, especially anisotropy.  In the future, similar
techniques combining self-assembly and DCM may become a powerful means
for assessing the magnetic properties of other nanoparticles, which
are too small to investigate individually.
\appendix
\section{Model for DCM response}
\label{app:model}
We follow Ref. \citenum{garcia-palacios_statics_2007} to establish a
model for the DCM response of an individual, thermally activated
Stoner-Wohlfarth particle.  We consider a particle with saturation
magnetization $M_s$, volume $V$, uniaxial anisotropy constant
$K_{u1}$, and cubic anisotropy constant $K_{c1}$.  The magnetization
vector of the macrospin is given by
$\mathbf{M}=M_s\cdot \mathbf{\hat{m}}$, where $\mathbf{\hat{m}}$ is a
unit vector.  The unit vector defining the uniaxial anisotropy is
given by $\mathbf{\hat{u}}$ and the unit vectors defining the cubic
anisotropy are $\mathbf{\hat{u}_i}$ with $i = 1, 2, 3$.  For
simplicity, we do not consider higher order anisotropies.  The
Hamiltonian of the magnetic system is then given by
$\mathcal{H} = \mathcal{H}_{\text{Zeeman}} +
\mathcal{H}_{\text{uniaxial}} + \mathcal{H}_{\text{cubic}}$, where,
\begin{equation}
  \mathcal{H}_{\text{Zeeman}} =  -\mu_0 \mathbf{M}\cdot\mathbf{H} V,
\end{equation}
\begin{equation}
  \mathcal{H}_{\text{uniaxial}}  =  - K_{u1}V \left(\mathbf{\hat{m}} \cdot \mathbf{\hat{u}}\right)^2,
\end{equation}
\begin{equation*}
\begin{split}
  \mathcal{H}_{\text{cubic}} = -V K_{c1} & \Big(\left(\mathbf{\hat{m}}\cdot\mathbf{\hat{u}_1}\right)^2 \left(\mathbf{\hat{m}}\cdot\mathbf{\hat{u}_2}\right)^2\Big.\\
  &  + \left(\mathbf{\hat{m}}\cdot\mathbf{\hat{u}_2}\right)^2 \left(\mathbf{\hat{m}}\cdot\mathbf{\hat{u}_3}\right)^2\\
  & \Big.+ \left(\mathbf{\hat{m}}\cdot\mathbf{\hat{u}_3}\right)^2
    \left(\mathbf{\hat{m}}\cdot\mathbf{\hat{u}_1}\right)^2\Big).
\end{split}
\end{equation*}

By substituting $-\frac{\mu_0}{2} D_u M_s^2$ for $K_{u1}$, the
uniaxial anisotropy can be expressed as a shape anisotropy, where
$D_u$ is the effective demagnetization
factor~\cite{aharoni_introduction_2000, skomski_permanent_1999}.  We
incorporate oscillations of the cantilever, which correspond to
rotations of $\mathbf{\hat{u}}$ and $\mathbf{\hat{u}_i}$ about
$\mathbf{\hat{y}}$ by an oscillation angle $\theta_c$, by applying a
rotation matrix $\mathbf{\widetilde{R}}(\theta_c)$ to all
$\mathbf{\hat{u}}$ and $\mathbf{\hat{u}_i}$.  The partition function
of the system is given by:
\begin{equation}
  \mathcal{Z} = \frac{1}{2\pi} \int^\pi_0 \diff\theta_m \sin\theta_m
\int^{2\pi}_0
\diff\phi_m\exp\left[-\frac{\mathcal{H}\left(\theta_m,\phi_m\right)}{k_B
    T}\right],
\end{equation}
where $\theta_m$ and $\phi_m$ are the polar and azimuthal angles of
$\mathbf{\hat{m}}$.  This yields the free energy, through
$\mathcal{F} = -k_bT\ln\mathcal{Z}$.  Once all parameter values are
set, the integral over $\theta_m$ and $\phi_m$ can be evaluated
numerically.  Using the difference quotient to approximate the second
derivative, we then calculate the frequency shift of the cantilever,
\begin{equation}
  \Delta f = \frac{f_0}{2k_0l_e^2}\left(\frac{\partial^2\mathcal{F}}{\partial \theta_c^2}\right) \approx \frac{f_0}{2k_0l_e^2}\frac{\mathcal{F}\left(\theta_c\right)-2\mathcal{F}\left(0\right)+\mathcal{F}\left(-\theta_c\right)}{\theta_c^2}.
\label{eq:finitediff}
\end{equation}

For the model fits shown in Fig.~\ref{fig:HT2}, we use a temperature of $T = \SI{270}{\kelvin}$, a saturation magnetization of $M_s = 3 \times 10^5\si{\ampere/\meter}$, a cubic anisotropy constant of $K_{c1} = \SI{-3.0}{\kilo\joule/\meter^3}$ and a volume of the individual nanoparticles of $(9\si{\nano\meter})^3$. The particle number is taken to be $n = 1.87 \times 10^6$, $0.83\times 10^6$ and $1.67\times 10^6$ and the uniaxial anisotropy constant $K_{u1} = 9.7$, $20.8$, and $\SI{2.1}{\kilo\joule/\meter^3}$ (or $D_u =-0.17$, $-0.37$, and $-0.04$) for samples 1,2 and 3, respectively.
The cantilever properties are $k_0 = \SI{314}{\micro\newton/\meter}$ and $l_{e} = \SI{74}{\micro\meter}$, and the angular oscillation amplitude is $\theta_c = 1.5^\circ$.

\section{Modeling a superparamagnetic particle}
\label{app:modelex}
\begin{figure}[t]
\includegraphics[width=8.49cm]{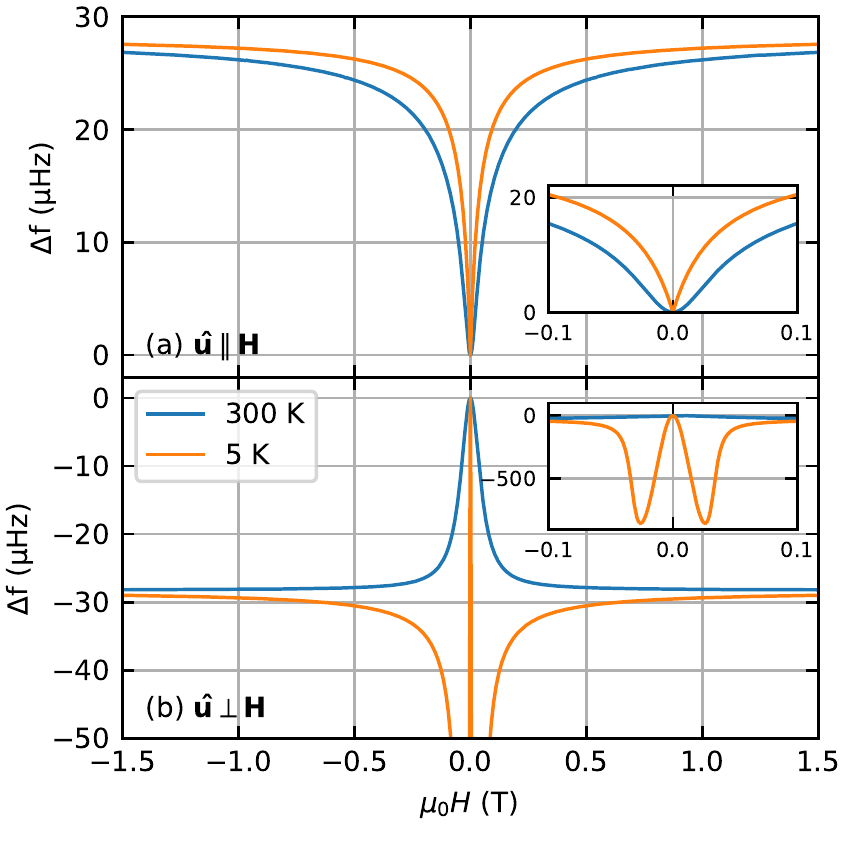}
\caption{Simulation of DCM for an individual, thermally-activated
  Stoner-Wohlfarth particle at $T = 300$ and $\SI{5}{\kelvin}$.
  Simulated $\Delta f(H)$ with (a)
  $\mathbf{\hat{u}} \parallel \mathbf{H}$, and (b)
  $\mathbf{\hat{u}} \bot \mathbf{H}$.}
\label{fig:suppl1}
\end{figure}

As an example of the DCM model discussed in Appendix~\ref{app:model},
we calculate $\Delta f$ of a magnetic particle for two different
temperatures, one at $T = \SI{5}{\kelvin}$ leading to a blocked state
and the other one at $T = \SI{300}{\kelvin}$ leading to a
superparamagnetic state.  The particle has a magnetic shape
anisotropy, given by the effective demagnetization factor $D_u$.  We
do each calculation for two different orientations of
$\mathbf{\hat{u}}$ and $\mathbf{H}$:
$\mathbf{\hat{u}} \bot \mathbf{H}$ and
$\mathbf{\hat{u}} \parallel \mathbf{H}$, as shown in
Fig.~\ref{fig:suppl1}.  We use the following parameters:
$f_0=\SI{5}{\kilo\hertz}$, $l_e = \SI{100}{\micro\meter}$,
$k_0 = \SI{100}{\micro\newton/\meter}$, $\theta_c = \SI{1}{\degree}$,
$V = \SI{1000}{\nano\meter^3}$, $M_s = \SI{300}{\kilo\ampere/\meter}$,
$D_u = -0.1$, $\theta_h = \phi_u = 0$, and
$\theta_u = \SI{0}{\degree}$, and $\SI{90}{\degree}$.

In the $\mathbf{\hat{u}} \parallel \mathbf{H}$ configuration, the
difference between the curves at $\SI{300}{\kelvin}$ and
$\SI{5}{\kelvin}$ is small: the curve at $\SI{300}{\kelvin}$ is
broader and approaches the horizontal asymptote more slowly than the
curve at $\SI{5}{\kelvin}$.  In the $\mathbf{\hat{u}} \bot \mathbf{H}$
configuration, however, the curves behave in a fundamentally different
way. The curve at $\SI{300}{\kelvin}$ is similar to the case of
$\mathbf{\hat{u}} \parallel \mathbf{H}$, but mirrored across
$\Delta f = 0$. The curve at $\SI{5}{\kelvin}$ has a distinct W-shape
for low fields, cf.\ the inset, approaching the horizontal asymptote
from negative rather than positive values of $\Delta f$.  This curve
matches the DCM curves calculated from the ferromagnetic
Stoner-Wohlfarth model of Ref.~\citenum{gross_dynamic_2016}. There is
a difference at low fields, where $\Delta f$ becomes positive in
Ref.~\citenum{gross_dynamic_2016}, but not in the model here. We
ascribe this to the different approaches to the problem: here, we have
a statistical model, that considers thermal excitation of the
magnetization. In Ref.~\citenum{gross_dynamic_2016}, a direct energy
minimization leads to the equilibrium magnetization and temperature is
not considered.  In the $\mathbf{\hat{u}} \perp \mathbf{H}$ case, the
distinction between the \reflectbox{\rotatebox[origin=c]{180}{V}}- and
W-shape of the DCM curve can be used to identify the para- or
ferromagnetic state of a magnetic specimen.

\begin{figure}[t]
\includegraphics[width=8.39cm]{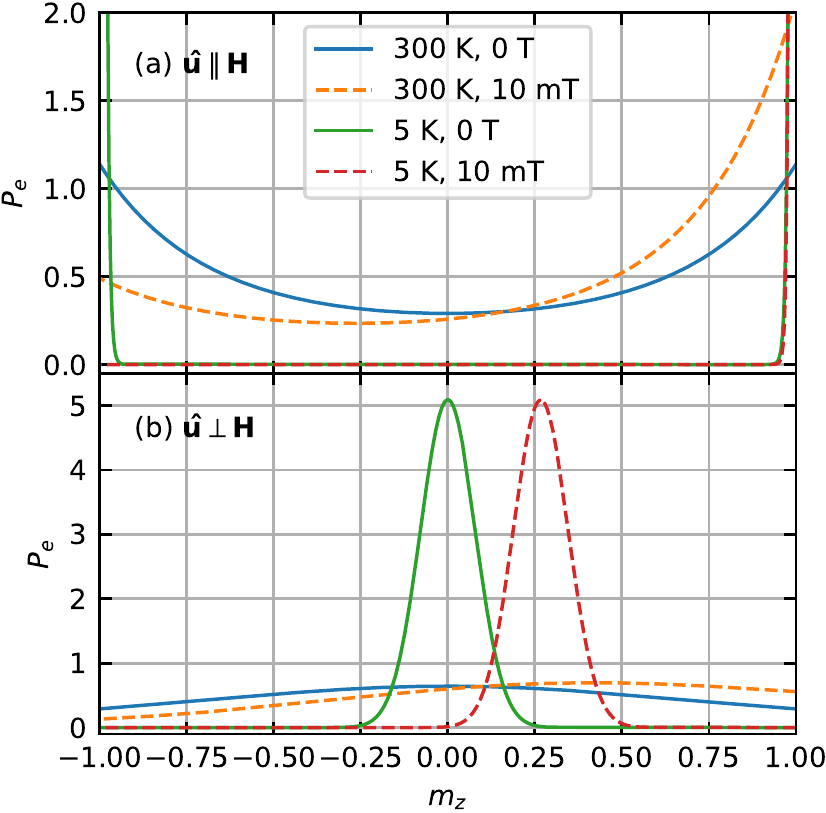}
\caption{$P_e$ for an individual, thermally activated Stoner-Wohlfarth
  particle with (a) $\mathbf{u} \parallel \mathbf{H}$, and (b)
  $\mathbf{u} \bot \mathbf{H}$.}
\label{fig:suppl2}
\end{figure}

To understand the progression of $\mathbf{M}$ with external field, it
is instructive to look at the equilibrium probability distribution of
magnetic moments, which is given by
\begin{equation}
  P_e\left(\theta_m,\phi_m\right) = \frac{\exp\left[-\mathcal{H}\left(\theta_m, \phi_m\right)/k_bT\right]}{\mathcal{Z}}.
\end{equation}
Integrating $P_e$ over $\phi_m$ and plotting it as a function of
$m_z = M_z/M_s = \cos\left(\theta_m\right)$ for a few values of the
external field, illustrates the difference between the blocked and the
superparamagnetic state, cf.\ Fig.~\ref{fig:suppl2}.  In (a), we plot
the probability for $\mathbf{\hat{u}} \parallel \mathbf{H}$ and
$H = 0$. For the ferromagnet at $\SI{5}{\kelvin}$ (green), there are
very sharp maxima at $m_z =\pm 1$. This means that $\mathbf{M}$ favors
alignment with $\mathbf{\hat{u}}$, with parallel and anti-parallel
alignment being equally probable. Away from these two peaks, $P_e$ is
essentially zero, i.e.\ the magnetization is very unlikely to point in
any direction other than along $\mathbf{\hat{u}}$. Turning on a slight
external field ($\mu_0H = \SI{10}{\milli\tesla}$), the peak at
$m_z = -1$ vanishes, and only $m_z = +1$ is favored (red
curve). Although not obvious in the DCM curves, $P_e$ shows how this
behavior differs from the paramagnetic behavior at $\SI{300}{\kelvin}$
(blue and orange curves). Here,
$\mathbf{M} \parallel \mathbf{\hat{u}}$ is still favored, but the
probability for $\mathbf{M}$ to point in any other direction is not
negligible. Turning on a small external field has a much smaller
effect on $P_e$ at $\SI{300}{\kelvin}$ than at $\SI{5}{\kelvin}$,
which explains why the DCM curve for the paramagnetic case is broader
than the ferromagnetic one.  Similar effects can be observed for
$\mathbf{\hat{u}} \bot \mathbf{H}$, cf.\ Fig.~\ref{fig:suppl2} (b). In
general, when an external field is applied, $\mathbf{M}$ rotates from
being aligned with $\mathbf{\hat{u}}$ towards the direction of
$\mathbf{H}$. For the ferromagnetic case, this is manifested in
relatively sharp peaks that shift towards $m_z = +1$ with increasing
field. For the paramagnetic case, very broad probability distributions
are present and the macro spin may fluctuate significantly in a broad
range of directions.

\section{Micromagnetic simulations}
Finite difference simulations are all performed with the Mumax software package~\cite{vansteenkiste_design_2014} using the following parameters, which are determined by the properties of the investigated samples and measurement apparatus:
$M_s = 3 \times 10^5$~\si{\ampere/\meter},
$A_{ex} = 10$~\si{\pico\joule/\meter\cubic},
$f_0 = 5572.134$~\si{\hertz}, $k_0 = 314$~\si{\micro\newton/\meter},
$l_{e} = 73.4$~\si{\micro\meter} and $\mu_0 H = 3.5$~\si{\tesla}. 
We minimize the energy of a magnetic state using a steepest descent method~\cite{exl_labontes_2014}, and then calculate the frequency shift using eq. (\ref{eq:finitediff}).
\subsection{Cubic shape anisotropy}
\label{app:cubic}
\begin{figure}[t]
\includegraphics[width=7.26cm]{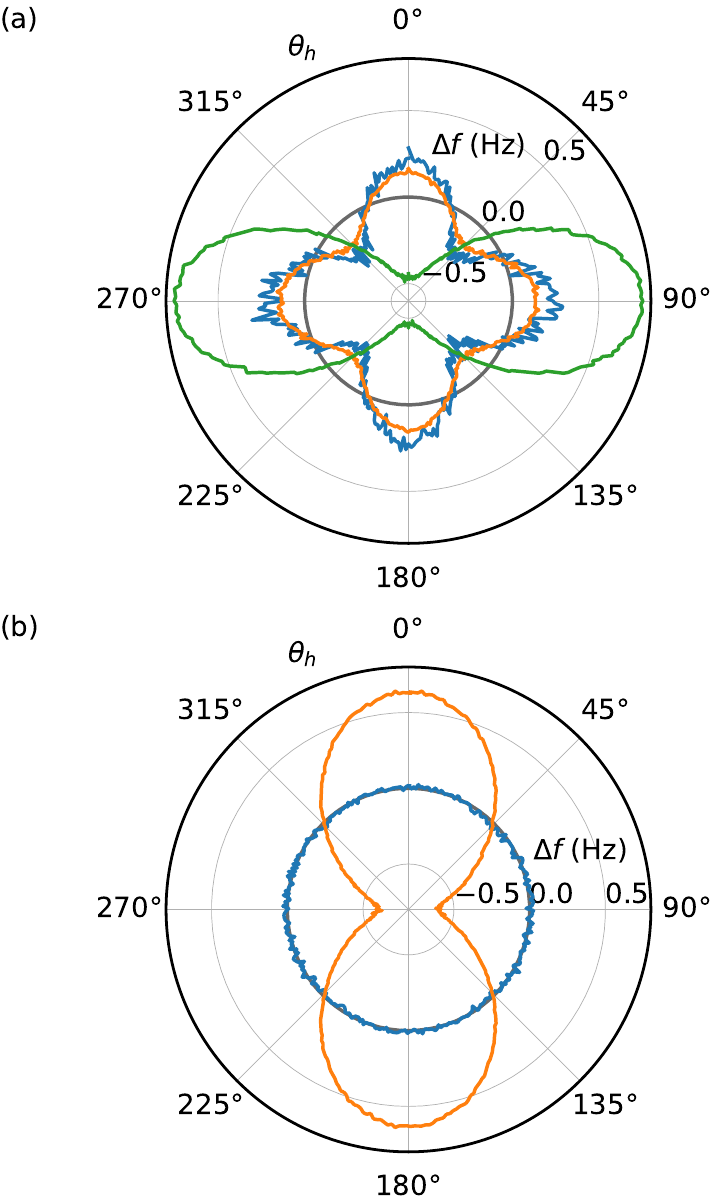}
\caption{(a) Simulated frequency shift of an individual nanoparticle of \SI{11}{\nano\meter} side length multiplied by $2\times 10^6$ (blue curve), a symmetric \SI{1}{\micro\meter} sized cube (orange curve) and a slightly asymmetric \SI{1}{\micro\meter} sized cube (green curve). (b) Simulated frequency shift of a BCT lattice of rounded nanocubes with lattice constants $a = c = \SI{14}{\nano\meter}$ (blue curve) and $a = \SI{14}{\nano\meter}$, $c = \SI{16}{\nano\meter}$ (orange curve).}
\label{fig:micromag}
\end{figure}

We analyze the symmetry and magnitude of the shape anisotropy of an individual maghemite nanoparticle by calculating $\Delta f(\theta_h)$ for a \SI{11}{\nano\meter} cube. In the simulations, the cube is discretized in cells with \SI{0.2}{\nano\meter} edge length. 
To estimate the impact of a single cube's anisotropy on $\Delta f$ of a full mesocrystal, we multiply with the approximate particle number $n\approx 2\times 10^6$ in a mesocrystal.
%
%
The simulation results are shown in Fig.~\ref{fig:micromag} (a) in blue. The symmetry of $\Delta f$ is cubic, as expected, and the magnitude is below \SI{0.5}{\hertz} for $\mu_0H = 3.5$~\si{\tesla}. This is far too small to explain the observed magnitude of the cubic component of $\Delta f$ (in the tens of \si{\hertz}) in experiment. To contrast this result, we add a cubic crystalline anisotropy with $K_{c1}= -\SI{3}{\kilo\joule/\meter\cubic}$ to the simulation and get a magnitude of $\Delta f$ of around \SI{25}{\hertz}, which is on the scale of the experimental results. Note that the real particles are rounded cubes as compared to a perfectly shaped cube in the simulation, further reducing the cubic shape anisotropy.

The same modeling procedure is carried out for a perfectly shaped $1\times 1 \times \SI{1}{\micro\meter\cubic}$ cube, in order to estimate the contribution of the mesocrystal's overall shape to the cubic component of the observed $\Delta f$. We use a cell size of \SI{10}{\nano\meter}, which is well below the exchange length of \SI{13.3}{\nano\meter} for the used material parameters, and have checked with \SI{5}{\nano\meter} cells that the results are robust against a further reduction of the cell size. The simulation results are shown in Fig.~\ref{fig:micromag} (a) in orange. Again, the magnitude of this effect is too small to account for the observed magnitude of the cubic component of $\Delta f$. For comparison, we show the result for a slightly asymmetric cube ($1\times 1 \times \SI{0.99}{\micro\meter\cubic}$) in green. The small asymmetry leads to a strong uniaxial component in $\Delta f$ as compared to the cubic component.
\subsection{Anisotropy due to the body centered tetragonal superlattice}
\label{app:BCT}
The periodicity of a mesocrystal's superlattice structure affects the effective shape anisotropy (see main text for definition) of the mesocrystal. Depending on the lattice constants, certain directions may have stronger or less strong dipolar interactions. To quantify
this influence, we carry out micromagnetic simulations of BCT lattices
of rounded cubes with different lattice constants. To avoid uniaxial
contributions to the shape anisotropy, resulting from elongation of
the mesocrystal in one direction, we choose a cubic simulation volume
of \SI{112}{\nano\meter} on a side. The cubes themselves have a side length of \SI{10}{\nano\meter} and rounded edges by intersecting with a \SI{12}{\nano\meter} sphere. We use periodic boundary conditions in all directions with 4 repetitions on each side of the simulation volume, so that edge effects are negligible (half- and quartercubes sitting on the edges and corners of the simulation volume guarantee correct periodicity). In this way, each spatial dimension of the simulated mesocrystal is equally sized with approximately \SI{1}{\micro\meter} length. Two sets of lattice constants are chosen, set 1 is $a = c = \SI{14}{\nano\meter}$ and set 2 $a = \SI{14}{\nano\meter}$ and $c = \SI{16}{\nano\meter}$, where $c$ points in $z$-direction in the coordinate system shown in Fig.~1. The calculated $\Delta f(\theta_h)$ for set 1, which is perfectly symmetric with respect to all three spatial dimensions, is shown as a blue curve in Fig.~\ref{fig:micromag} (b). $\Delta f$ is negligible in all directions compared to the value measured in our experiment. Set 2, for which the cubes have a slightly larger spacing in $z$-direction, shows an easy uniaxial ansiotropy contribution in
this direction. However, the magnitude of the effect is around $\SI{0.5}{\hertz}$, and hence small compared to the effect of elongations of the mesocrystal in one spatial direction. The lattice constants of set 2 and their difference are very comparable with the values of the experimentally investigated samples ($a = \SI{13.47}{\nano\meter}$ and $c = \SI{15.08}{\nano\meter}$), and we thus expect a minor contribution of the superlattice structure on the shape anisotropy of the samples.
\section{Measurement of the superparamagnetic blocking temperature}
\label{app:blockT}
\begin{figure}[t]
\includegraphics[width=8.22cm]{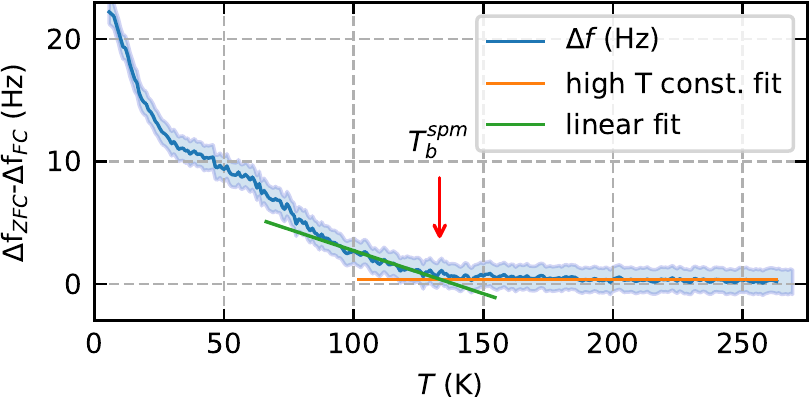}
\caption{Field cooling data for sample 3. The graph shows the
  difference between the frequency shift for a zero field cooling and
  a field cooling measurement in $\mu_0 H = \SI{125}{\milli\tesla}$.}
\label{fig:TD_fc}
\end{figure}

Typically, the blocking temperature $T_b^{\text{spm}}$ of a
superparamagnetic system is identified by comparing FC
to ZFC magnetization
measurements~\cite{bedanta_supermagnetism_2008}. DCM does not give
access to the magnetization
($M \propto \partial \mathcal{F} / \partial H$), but rather to the
curvature of $\mathcal{F}$ with respect to $\theta_c$
($\Delta f \propto \partial^2 \mathcal{F} / \partial \theta_c^2$).
$\Delta f$ is hence comparable to the magnetic susceptibility
($\chi \propto d^2\mathcal{F}/dH^2$).  Frequency dependent
measurements of $\chi$ allow the identification of
$T_b^{\text{spm}}$~\cite{bedanta_supermagnetism_2008}.  Frequency
dependent DCM measurements, however, are complicated by the
cantilever's discrete mechanical modes.  We can, however, compare FC
and ZFC measurements of $\Delta f$.  Fig.~\ref{fig:TD_fc} shows the
difference between these frequency shifts,
$\Delta f_{\text{ZFC}} - \Delta f_{\text{FC}}$, for sample~3.
The mesocrystal is first cooled in zero field; then
$\Delta f_{\text{ZFC}}$ is recorded while the sample is heated to room
temperature with $\SI{125}{\milli\tesla}$ of field applied.  To
measure $\Delta f_{\text{FC}}$, the same procedure is repeated, but
with field-cooling in $\SI{1}{\tesla}$.  For temperatures above
$\SI{133}{\kelvin}$ there is no difference between ZFC and FC
measurements. Below this temperature, a difference begins to appear,
suggesting that the individual nanoparticles stop behaving like
superparamagnets and begin behaving like ferromagnets, i.e.\ they are
blocked.  Therefore, we conclude that the blocking temperature is
$T_b^{\text{spm}} \approx\SI{133}{K}$ for these mesocrystals.

This temperature compares well with data from ensemble measurements of
mesocrystals with similar sizes \cite{wetterskog_tuning_2016}
($\SI{125}{\kelvin}$ for $\SI{9.6}{\nano\meter}$ and
$\SI{155}{\kelvin}$ for $\SI{12.6}{\nano\meter}$ particles, while the
present ones are $\SI{10.9}{\nano\meter}$ sized). Measurements on a
dilute ensemble of similar-sized maghemite particles size, but with
large silica shells to suppress inter-particle interactions, show
$T_b^{\text{spm}}
\approx\SI{60}{K}$\cite{andersson_interacting_2017}. Hence, the
interactions between the particles in the mesocrystal appear to
increase $T_b^{\text{spm}}$ significantly.

\begin{figure}[t]
\includegraphics[width=8.5cm]{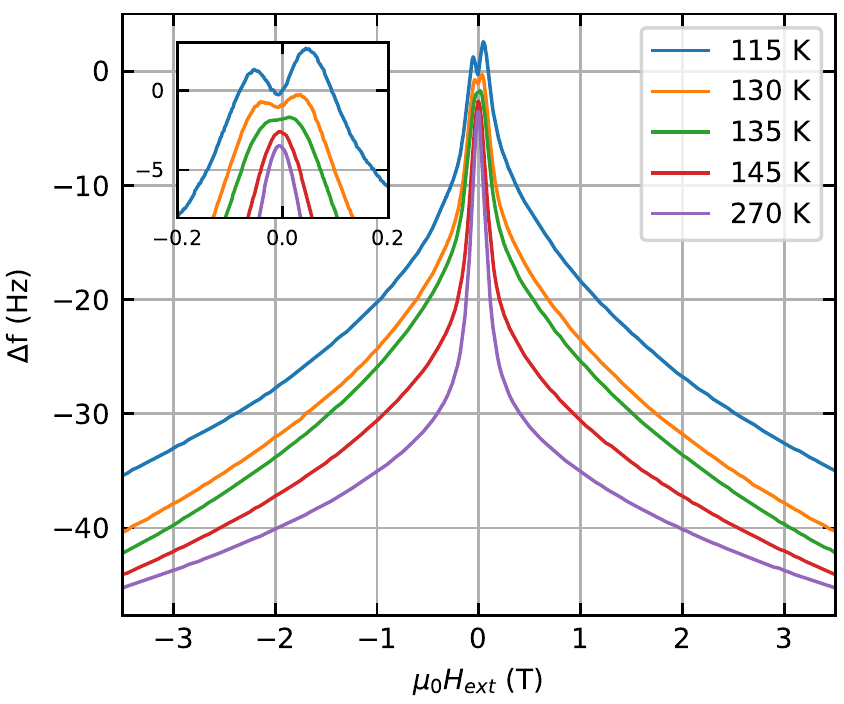}
\caption{DCM response vs.\ external field with
  $\theta_h=\SI{180}{\degree}$ for various temperatures around the
  blocking temperature for sample 2. The curves are offset for better
  visibility.}
\label{fig:TD1}
\end{figure}

Temperature dependent hysteresis data, recorded for sample~2, provide
further support for the value of $T_b^{\text{spm}}$,
cf. Fig.~\ref{fig:TD1}.
With the applied magnetic field aligned along the hard axis, the shape
of $\Delta f(H)$ drastically changes at $T_b^{\text{spm}}$. Above
$T_b^{\text{spm}}$, the data match the predictions of our model of
interacting superparamagnets. Below $T_b^{\text{spm}}$, however, the
maximum in $\Delta f$ at zero field transforms into a asymmetric
\reflectbox{\rotatebox[origin=c]{180}{W}}-shape with two maxima.  From
the shape of the DCM curves predicted by our model for the para- and
ferromagnetic states, we can identify the low-temperature onset of
ferromagnetism.

\section{Behavior below the superparamagnetic blocking temperature}
\label{app:belowBlock}

\begin{figure}[t]
\includegraphics[width=8.06cm]{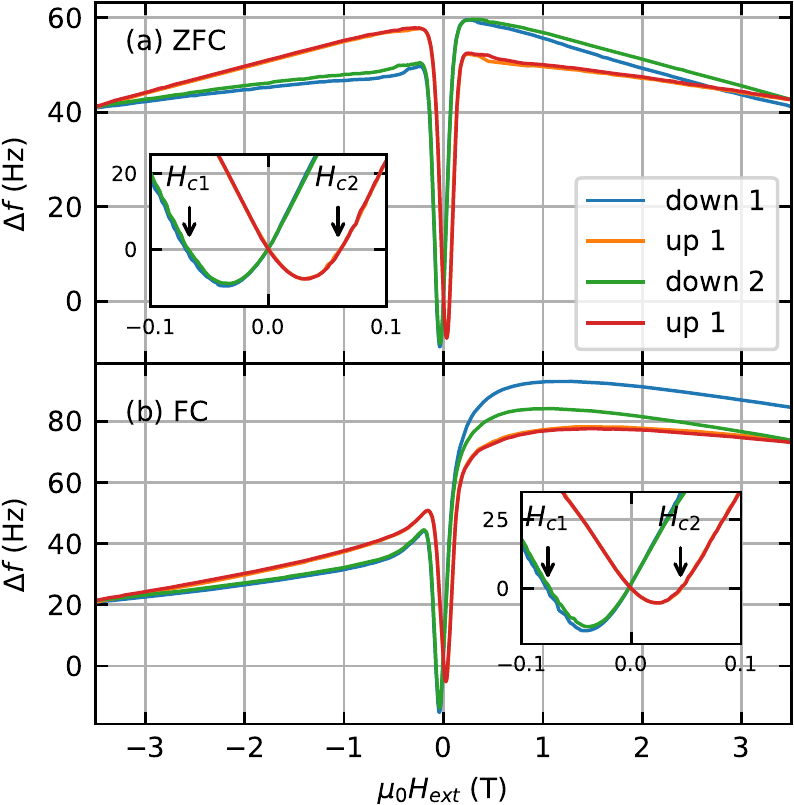}
\caption{Two consecutive $\Delta f$ hysteresis loops at
  $\SI{5}{\kelvin}$ for (a) ZFC and (b) FC measurements with
  $\theta_h = \SI{184}{\degree}$ for sample 3.}
\label{fig:TD2}
\end{figure}

Hysteresis loops taken at $T = \SI{5}{\kelvin}$ for sample 3 show that
at low temperatures not only superparamagnetism is blocked, but a more
complicated magnetic state is present. Depending on the cooling
procedure, the measurement proceeds differently, as can be seen in
Fig.~\ref{fig:TD2} for measurements with (a) ZFC
and (b) FC in \SI{3.5}{\tesla}.
Three main observations can be made from the low temperature
measurements: First, the hysteresis loops do not saturate even for the
highest applied fields. Second, there is exchange bias
\cite{phan_exchange_2016} present, which can be seen by comparing the
positive and negative coercive fields ($\Delta f = 0$) in the insets
of the figure. Both statements are true irrespective of the cooling
procedure. Third, we find a highly asymmetric behavior with respect to
the sign of the external field for the FC measurement. The high-field
frequency shift differs strongly for positive and negative field
values, which is considerably reduced for a second consecutive
hysteresis loop.  This means that the system can be trained, a typical
behavior in a system with an exchange bias.  To further understand
these findings, we analyze temperature dependent measurements of the
coercive field and exchange field.

\begin{figure}[t]
\includegraphics[width=8.3cm]{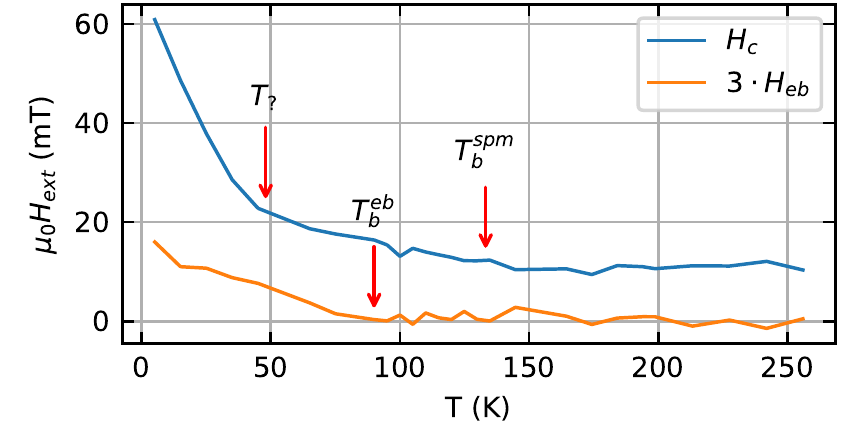}
\caption{$H_c$ and $H_{\text{eb}}$ vs $T$
  for 
  zero field cooling in easy anisotropy orientation for sample 2 with
  $\theta_h = \SI{180}{\degree}$.}
\label{fig:TD3}
\end{figure}

The blue line in Fig.~\ref{fig:TD3} shows the temperature dependence
of the coercive field $H_c = \left|H_{c2}-H_{c1}\right|/2$, where
$H_{c1}$ ($H_{c2}$) is the negative (positive) coercive field for a
ZFC measurement.
The coercivity $H_c \approx \SI{10}{\milli\tesla}$ is constant above
$T_b^{\text{spm}}$, as we may expect for a superparamagnetic system
with strong interactions. Below $T_b^{\text{spm}}$, $H_c$ starts to
increase, suggesting that the net magnetic moments of an increasing
number of nanoparticles switch collectively with decreasing
temperature. Just below $\SI{50}{\kelvin}$ the curve steepens
significantly.  This may indicate a magnetic transition of unknown
origin.

$H_{c1}$ and $H_{c2}$ show the same magnitude above
$\SI{90}{\kelvin}$. Below \SI{90}{\kelvin}, $H_{c1}$ becomes larger
than $H_{c2}$ in magnitude. This effect can be quantified by the
exchange bias field $H_{\text{eb}} = \left|H_{c2}+H_{c1}\right|/2$ and
is shown as orange curve in Fig.~\ref{fig:TD3}. From this data, we
infer a blocking temperature of the exchange bias effect of
$T_{b}^{\text{eb}}\approx\SI{90}{\kelvin}$. Below $T_{b}^{\text{eb}}$,
$H_{\text{eb}}$ increases moderately with decreasing
temperature. Doing the same experiment after a FC procedure leads to a
significantly enhanced $H_{eb}$.
\begin{acknowledgments}
  We thank Sascha Martin and his team in the machine shop of the
  Physics Department at the University of Basel for help building the
  measurement system.  We acknowledge the support of the Canton Aargau
  and the Swiss National Science Foundation under Project Grant
  200020-159893, via the Sinergia Grant Nanoskyrmionics (Grant
  No. CRSII5-171003), and via the NCCR Quantum Science and Technology
  (QSIT).  J. L. is supported by the Research Foundation - Flanders
  (FWO) through a postdoctoral fellowship. L.B. acknowledges the
  Swedish Research Council (VR, grant number: 2019-05624) for funding
  this work, and we gratefully acknowledge Doris Meertens for FIB
  preparation of the TEM lamella and Andr\'as Kov\'acs for performing
  the TEM image.
\end{acknowledgments}
%
%

%

\end{document}